%% file: masterfile.tex
\documentclass[10pt,twoside,BCOR7mm,DIV15,headinclude,footexclude,
               cleardoubleempty,idxtotoc]
{scrartcl}

\usepackage{natbib}
\usepackage[font=small,labelfont=bf]{caption}
\usepackage[english]{babel}
\usepackage{graphicx}
\usepackage{hyperref}
\usepackage{scrpage2}
\usepackage{ifthen}
\usepackage{booktabs}
\usepackage{amsmath}
\usepackage{amssymb}
\usepackage{multicol}
\usepackage{float}
\usepackage{hyperref}

\hypersetup{breaklinks=true
,colorlinks=true,linkcolor=black,urlcolor=blue
,citecolor=black}

\addto\captionsenglish{%
}

\makeatletter
\renewcommand{\@biblabel}[1]{}
\renewcommand{\@cite}[2]{%
{#1\ifthenelse{\boolean{@tempswa}}{,#2}{}}}
\makeatother
\ofoot{\thepage}
\ifoot{}

\setcapindent{0em}
\setheadsepline{1pt}

\setkomafont{pagehead}{\normalfont\sffamily}
\setkomafont{pagenumber}{\normalfont\rmfamily}

\makeatletter
\newcommand{\listofcontributions}{\@starttoc{con}}

\newcommand{\l@contribution} {\@dottedtocline{1}{1.5em}{2.3em}}
\makeatother

\newenvironment{contribution}{
\setcounter{section}{0}
\setcounter{figure}{0}
\setcounter{table}{0}
}{
\newpage
\lehead{}
\rohead{}
}

\begin{document}

\setlength{\baselineskip}{2.5ex}

\begin{contribution}
\include{myarticle}

\end{contribution}

%%-------------------------------------------------------

\end{document}

%% file: myarticle.tex
% EXAMPLE AND TEMPLATE FILE FOR PROCEEDINGS OF THE WOLF-RAYET WORKSHOP.
% PLEASE REPLACE THE TEMPLATE TEXT BY YOUR OWN ARTICLE.
% NOTE THAT YOU MUST NOT PROCESS THIS FILE, BUT THE MASTER FILE:
% latex masterfile; dvips masterfile

% RUNNING AUTHOR: PUT AUTHOR NAMED HERE
\lehead{G.\ Meynet, C.\ Georgy, A.\ Maeder, S.\ Ekstrom, J.\ Groh, H.F.\ Song \& P. \ Eggenberger}

% RUNNING TITLE; SHORTEN THE TITLE IF NECESSARY
% IN CASE OF A ONE-PAGE CONTRIBUTION (POSTER),
% SQUEEZE AUTHORS AND TITLE IN THIS LINE (Author: Title ...)
\rohead{Physics of stellar models}

\begin{center}
% FULL TITLE HEADING
{\LARGE \bf Physics of massive stars relevant for the modeling of Wolf-Rayet populations}\\
\medskip

% AUTHORS LIST
{\it\bf G.\ Meynet$^1$, C.\ Georgy$^{1,2}$, A.\ Maeder$^1$, S.\ Ekstrom$^1$, J.\ Groh$^1$, F.\ Barblan$^1$ \& H.F.\ Song$^3$ \& P. \ Eggenberger$^1$}\\

% AFFILIATIONS
{\it $^1$Geneva University, Switzerland}\\
{\it $^2$Keele University, UK}\\
{\it $^3$ Guizhou University, China}

% ABSTRACT
\begin{abstract}
Key physical ingredients governing the evolution of massive stars are mass losses, convection and mixing in radiative zones. These effects are important both in the frame of single and close binary evolution. The present paper addresses two points: 1) the differences between two families of rotating models, {\it i.e.} the family of models computed with and without an efficient transport of angular momentum in radiative zones; 2) The impact of the mass losses in single and in close binary models.
\end{abstract}
\end{center}

% TEXT OF THE PAPER, TWO-COLUMN STYLE
\begin{multicols}{2}

\section{Rotation}

\begin{figure}[H]
\begin{center}
\includegraphics[width=\columnwidth]{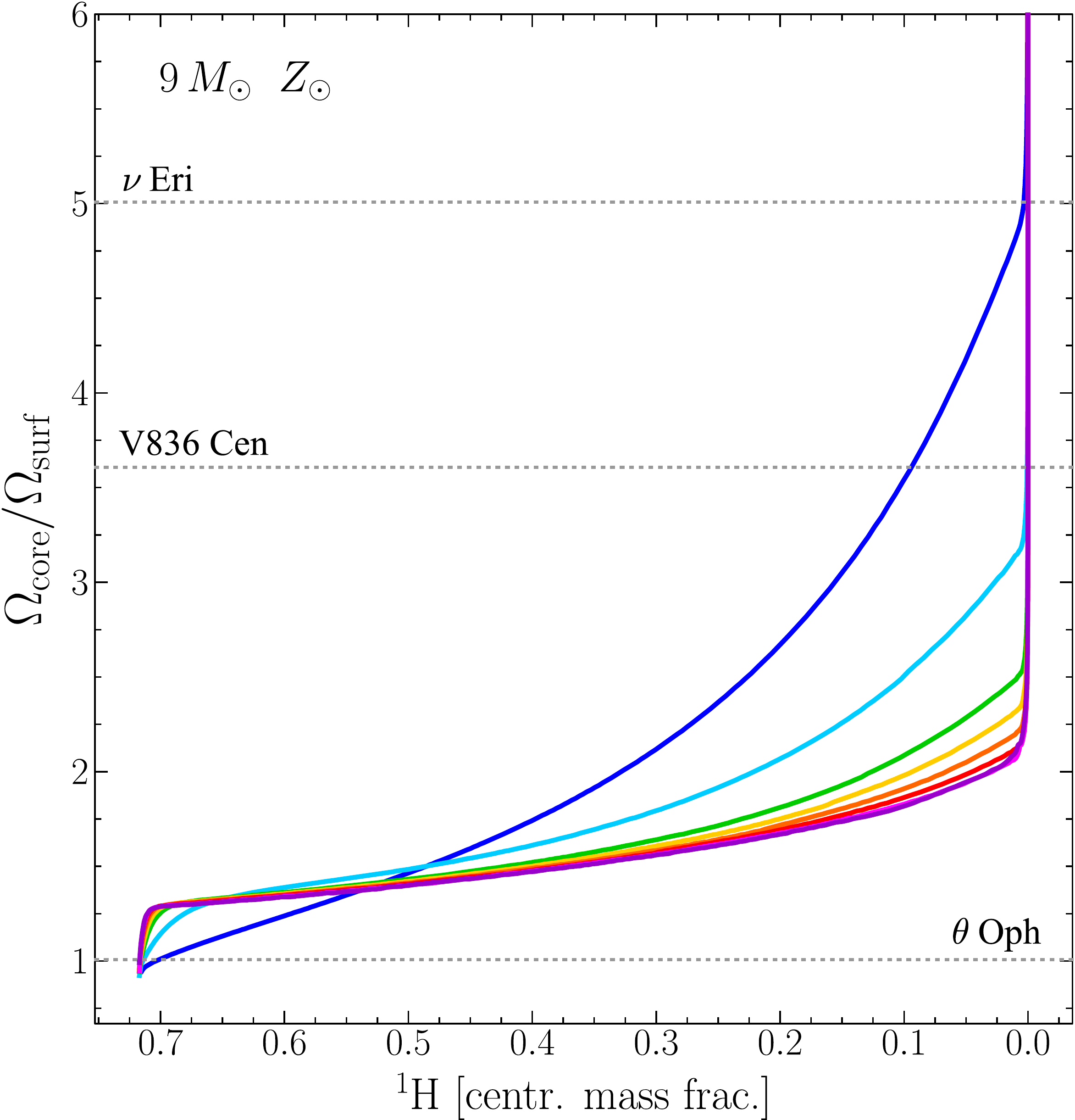} 
\caption{Variation of the ratio between the angular velocity of the core and the angular velocity at the surface as a function of the remaining mass fraction of hydrogen in the core during the core H-burning phase in 9 M$_\odot$ models for a metallicity $Z$=0.014. The lines from top to bottom correspond to increasing initial rotation on the ZAMS, they correspond to initial surface angular velocities equal to 0, 10, 30, 50, 60, 70, 80, 90 and 95\% the critical angular velocity (the critical angular velocity being the value
for which the centrifugal acceleration balances the gravity at the equator). The ratios estimated by asteroseismology are indicated for a few cases as horizontal dotted lines \citep[][and references therein]{Aerts2008}. The models are those by \citet{Georgy2013}.
\label{neuf}}
\end{center}
\end{figure}

Present extended grids of massive star rotating models  \citep[see e.g. the review by][and references therein]{MM2012} can be classified into two main families (note that this is true for both single and close binary evolution). The first family consists in models where the transport of the angular momentum is driven by meridional currents and shear instabilities according to the theory proposed by  \citet{Zahn1992} \citep[see for instance the grid by][] {ekstrom12}.  We shall call these models, the
shear models. The shear models allow a small contrast between the angular velocity of the convective core and the angular velocity of the surface  to develop during the Main-Sequence phase as shown in Fig.~\ref{neuf}. The contrast is more pronounced in models starting with a slow initial rotation. This is due to the fact that when the initial rotation is smaller, the processes transporting the angular momentum are slower. Thus the contrast in rotation between the core and the envelope, that builds up during the Main-Sequence phase when the core contracts and the envelope expands, is less smoothed. In Fig~\ref{neuf}, the contrast deduced from asteroseismology for three stars are indicated by horizontal dotted lines. We see that for two stars, the measured values are compatible with those of initially slowly rotating models, while for the third one, the observed result is compatible with solid body rotation or  with the models shown in Fig.~\ref{neuf} at a very early stage of the core H-burning phase. It is interesting to note that the good agreement with initially slowly rotating models is consistent with the fact that such asteroseismic estimates are only available for slow rotators. 

The second family is composed from models computed with the dynamo mechanism proposed by \citet{Spruit02}. Among grids of massive star models computed with that theory we can mention for instance
the grids by  \citet{Brott2011}. We shall call these models, the magnetic models.
This  theory  has been criticized by \citet{Zahn07}, but we 

shall not develop here the arguments, we rather 
point a few similarities/differences when the outputs are compared with those of the shear models. Among the similarities, we can mention that, in the two families of models, the efficiency of the chemical mixing varies qualitatively in the same way when the initial rotation, mass and metallicity vary. For both families, when the rotation increases, the mixing becomes more efficient, the same occurs when the initial mass increases; when the metallicity increases, the mixing becomes less efficient. These similarities are striking because they are due to different physical reasons in both families \citep[see a more detailed discussion in][]{Song2015}.

%-------- Table over one column -----------------------------------
%\begin{table}[H] 
%\begin{center} 
%\captionabove{$\Omega_{\rm core}/\Omega_{\rm env}$ estimated by asteroseismology \citep{Aerts2008, Goupil2011}} 
%\label{ast}
%\begin{tabular}{lc}
%\toprule
%star   & $\Omega_{\rm core}/\Omega_{\rm env}$  \\
%\midrule 
%$\nu$ Eri   & $\sim$5 \\
%V836 Cen & 3.6 \\
%$\theta$ Oph  & $\sim$1 \\
%12 Lac  & 1.8-5 \\
%\bottomrule
%\end{tabular}
%\end{center}
%\end{table}
%--------------------------------------------------------------------

\begin{figure}[H]
\begin{center}
\includegraphics[width=\columnwidth]{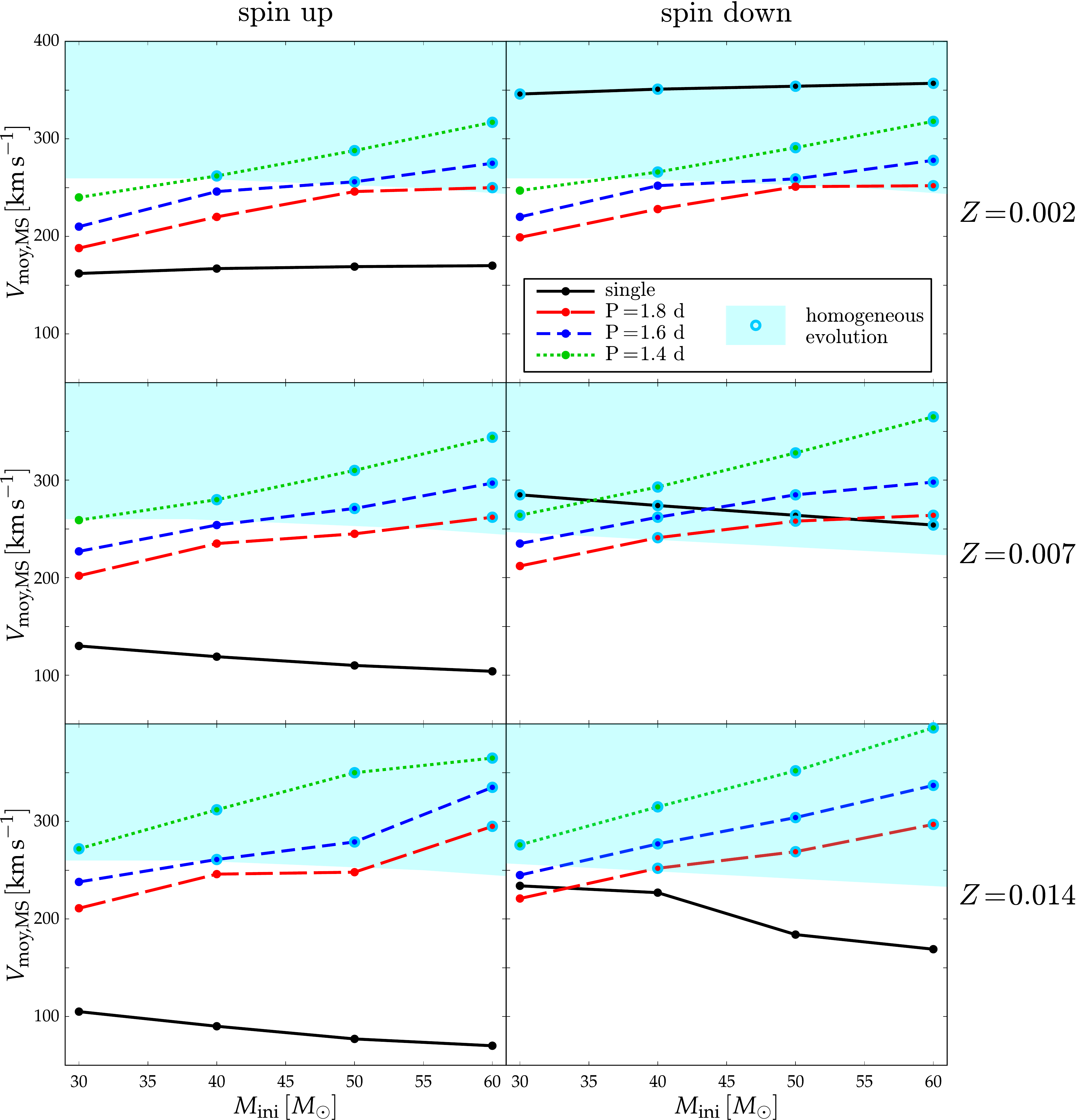}
\caption{Time-averaged surface equatorial velocity of single and binary stars as a function of the initial mass for various metallicities. The models have been computed with an internal magnetic field. The (black) continuous line connects the single star models. The (colored) dashed lines connect the binary stellar models.
 The initial period in days is indicated. The results corresponds to a companion having 2/3 the mass of the primary. The (blue) big dots show the models that follow a homogeneous evolution. The zone of homogeneous models lay  is
hatched in blue. The panels on the left column correspond to spin-down cases and the panels on the right column to spin-up cases for binaries. The figure is taken from \citet[][and see the text for more details on spin-up and spin-down cases.]{Song2015}.
\label{veq}}
\end{center}
\end{figure}

Let us now turn to the differences. A main difference is
that magnetic models produce nearly solid body rotating models during the Main-Sequence (MS) phase. If plotted in a figure like Fig.~\ref{neuf}, they would show quasi horizontal lines fixed at an ordinate equal to 1. Such model could be compatible with the case of $\theta$ Oph. A second difference is that the magnetic models, with no magnetic braking at the surface, all other characteristics being kept the same (mass, metallicity, initial rotation, age) are more efficiently mixed than shear models \citep{MMIII2005}. This implies that magnetic models will for instance produce a homogeneous evolution ({\it i.e.} an evolution during which the chemical composition at the centre of the star is quasi the same as the chemical abundance at the surface) with an initial rotation that is smaller than for shear models. Fig.~\ref{veq} show the time-averaged surface velocity for different magnetic models, single or in wide binaries (see the continuous black lines) and in close binaries (see the colored dashed lines), starting with different initial velocities. For these magnetic models, it suffices that the time-averaged surface velocity be above $\sim$250 km s$^{-1}$ to follow an homogeneous evolution, while for shear models, much larger velocities are needed (see below). This last point can be illustrated comparing the evolution of the magnetic 39 M$_\odot$ model computed with an initial rotation of 350 km s$^{-1}$ by \citet{Szecsi2015} for the metallicity of IZw18. This model follows a homogeneous evolution. The corresponding shear model 
(initial mass of 40 M$_\odot$) computed for an initial velocity of 393 km s$^{-1}$ follows a regular, non-homogeneous evolution (Groh et al. in preparation). Thus, the link between a given evolution (here for instance between homogeneous or non homogeneous evolution) and the initial rotation is thus very different depending on the type of models (shear and magnetic) considered.
For  a given initial distribution of velocities, the predicted outputs for a population of massive stars and in particular the populations of WR stars can therefore be significantly different. This has to be kept in mind
when comparisons with observations are done.

At the moment, it is not possible to make strong conclusions about which kind of models would be the most representative of the behavior of the bulk of the real stars. Some indirect arguments as the angular momentum content in white dwarfs and neutron stars support magnetic models \citep{Suijs2008, Heger2005}. The magnetic models allow to extract more angular momentum from the central 

%-----------One-column figure -----------------------------------
% Note that only the [H] option is allowed for placing 1-column figures!
%-----------------------------------------------------------

regions and thus predict final rotations for the compact remnants that are more compatible with the observations than
the predictions of the shear models that predict too high rotation for these remnants. On the other hand, as indicated above, magnetic models predict solid body rotation during the MS phase, which is in contradiction with some asteroseismic constraints. Is there any possibility to reconcile the asteroseismic constraints and those coming from the measured rotation rates of young pulsars? Let us make here a few remarks: 1) the rotation of young pulsars might result from processes occurring at the time of the supernova explosion, or during the early phases of the new born neutron stars; 2) recent works have suggested other mechanisms to extract angular momentum from the core, very different from the
theory by \citet{Spruit02}. One of them invokes a coupling between the core and the surrounding radiative layers by a fossil magnetic field \citep{MMIV2014}. \citet{Fuller2015} suggests another way by
demonstrating that internal gravity waves, excited via envelope convection during a red supergiant phase or during vigorous late time burning phases, may substantially spin down the core.

\begin{figure}[H]
\begin{center}
\includegraphics
  [width=\columnwidth]{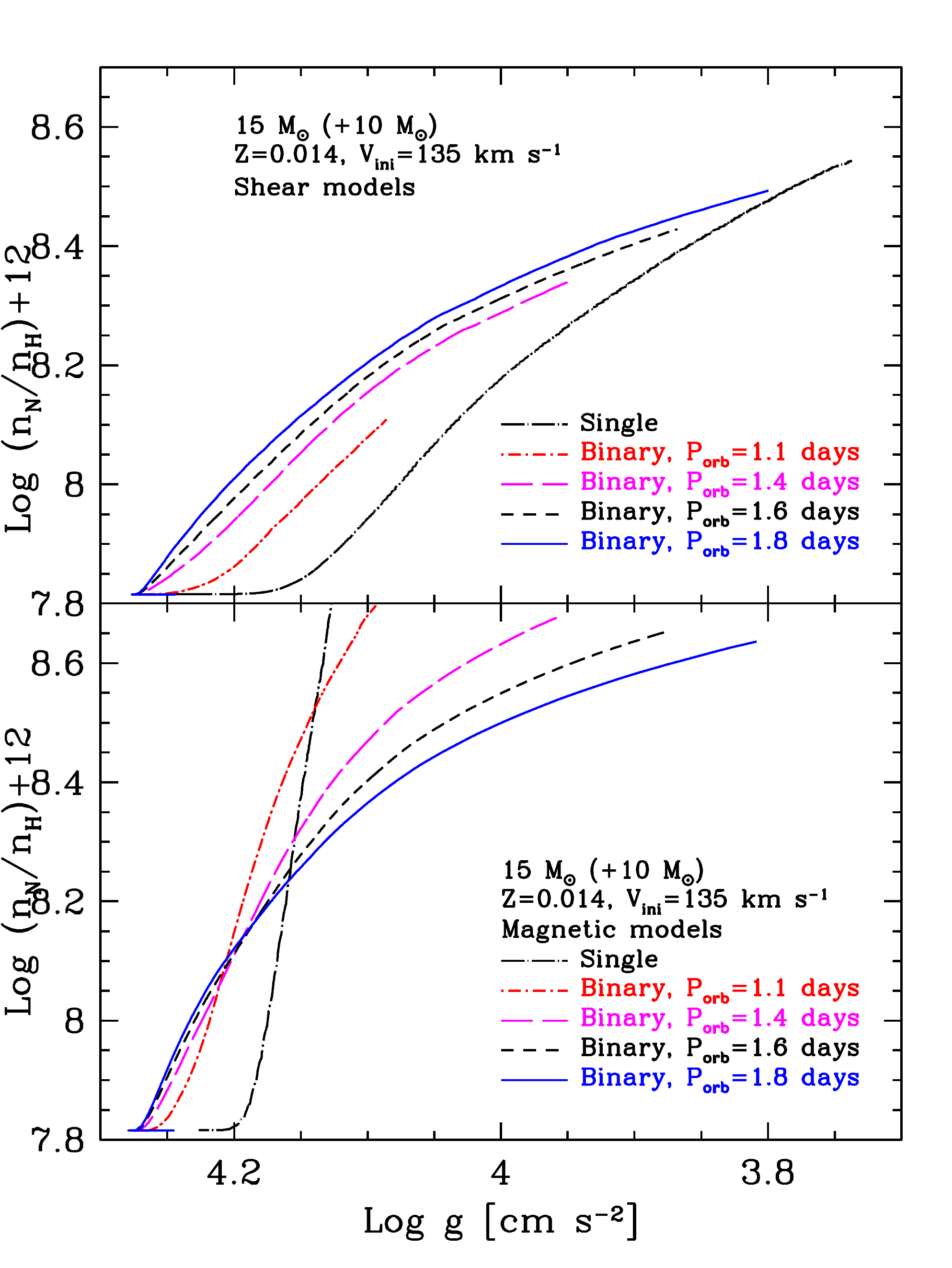}
\caption{Evolution of the abundance ratio (in number) of nitrogen to hydrogen at the surface of rotating 15 M$_\odot$ models at solar metallicity as a function of the surface gravity.  Single models and models in close binaries are shown. The initial velocity is 135 km s$^{-1}$.{\it Upper panel:}  the models have been computed according to the rules of the shear models. Figure taken from \citet{Song2013}. {\it Lower panel:} the models have been computed according to the rules of the magnetic models and are discussed in \citet{Song2015}.
\label{NHshear}}
\end{center}
\end{figure}

To make progresses, we need to find some additional discriminating observations that would allow to check which kind of models seem the most appropriate. It might be that
shear models and magnetic models both occur in nature. In that case it is important to understand what are the causes of these different behaviors and what are their respective frequency.
A possible way of differentiating these two kinds of models will be through the observations of stars in close binaries. As can be seen in the upper panel of Fig.~\ref{NHshear}, shear models in close binaries show larger surface enrichments when the orbital period increases, while we have the reverse behavior in magnetic models (see the lower panel Fig.~\ref{NHshear}). This comes from the fact that
in shear models, the mixing is driven by the gradients of $\Omega$, while in magnetic models, the mixing is driven by $\Omega$. In shear models, a larger orbital period imposes a slower rotation at the surface and thus a stronger contrast with the centre making the gradients of $\Omega$ stronger and the mixing stronger. In magnetic models, a larger orbital period imposes a lower value of $\Omega$ in the whole star and thus weakens the efficiency of mixing.

\section{Mass loss rates}

Mass loss rates by stellar winds and/or through mechanical winds when the star is rotating near the critical limit are also very important quantities relevant for WR star modeling.
Actually the intensity of the winds {\it prior} the star enters the WR regime has a strong impact on the duration of the WR lifetime and hence on estimating quantities as the
number ratio of WR to O-type stars. Higher the mass loss rates, longer will be the WR lifetime. This is illustrated for instance by looking at how, in the frame of single star models, the lifetime
of 60 M$_\odot$ models in the WR phase increases when the initial metallicity and therefore the mass loss rates increase \citep[see for instance Fig. 7 in][]{MMWR2005}.
On the other hand, mass loss rates {\it during} the WR phases 
may significantly change the durations of the various
WR subphases. For instance, increasing the mass loss rate during the WN phase would reduce the duration of the WN phase and increases the duration of the WC phase, having a strong impact on the number ratios of WC to WN stars.

In order to know whether a given model has to be considered a WN or a WC star, the most precise procedure would be to compute the emergent spectrum compatible with the
interior structure and then apply to this synthesized spectrum the same criteria of classifications, based on various line ratios, to assign a spectral WR subtype. At the moment, there is

%--------------------------------------------------------------------
%\begin{figure}[H]
%\begin{center}
%\includegraphics
%  [width=\columnwidth]{gncmagn.eps}
%\caption{Same as Fig.~\ref{NHshear}, but the models have been computed according to the rules of the magnetic models. The models are those of \citet{Song2015}
%\label{NHmagn}}
%\end{center}
%\end{figure}

%----------- Double-column figure -----------------------------------

only one complete evolution for which spectra were 
computed all along the stellar life. This was done for a non-rotating 60 M$_\odot$ model at solar metallicity by \citet{Groh2014}. It is interesting to compare the
lifetimes obtained for the different evolutionary phases when spectral criteria are used and when more global criteria based on the effective temperature and the surface abundances are used
\citep[for a detailed comparison see Table 3 in][]{Groh2014}. Some durations are the same whatever the spectral or the global properties are used. This is for instance the case for the total duration of the
WR phase which differs by only half a percent between the two methods. Others are strongly affected. For instance, the spectral classification gives a much shorter WNL phase (the spectral duration corresponds to 8\% the duration obtained by the global properties) and a much longer WNE phase
(the spectral duration corresponds to about 2 times the duration obtained by the global properties). This spectral modeling, we see would be here particularly important for predicting number ratios involving WNL and WNE stars and for
making comparisons with observed positions of this two types of stars in the HR diagram. Let us note that the mass loss rates used have an impact on the output spectra. More consistent models would be obtained if the mass loss rates would result from the model physics rather than from some empirical recipes.

\begin{figure}[H]
\begin{center}
 \includegraphics[width=\columnwidth]{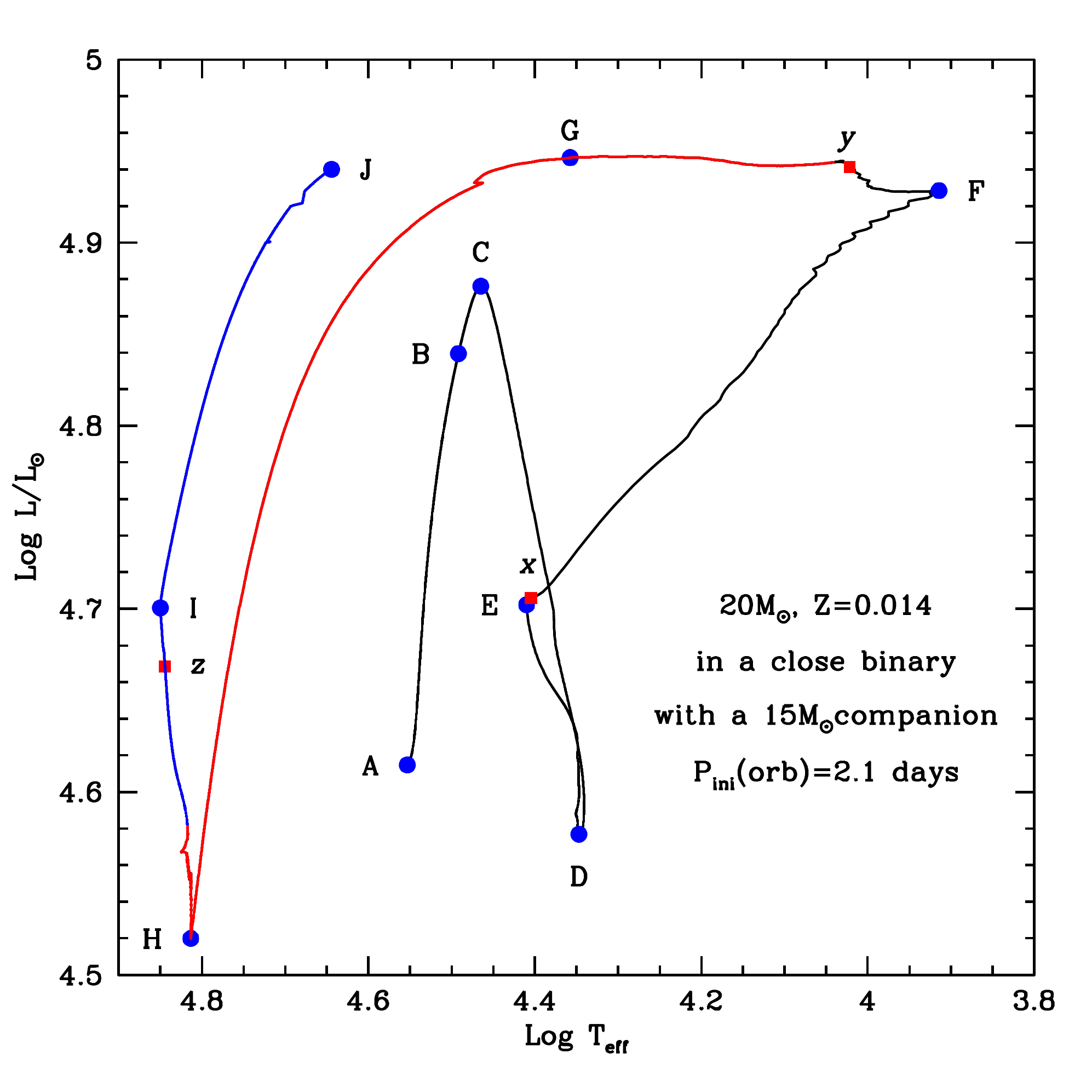}
\caption{Evolutionary track in the theoretical HR diagram for a primary of 20 M$_\odot$ in a close binary system with a 15 M$_\odot$ star. The initial orbital period
is 2.1 days. The metallicity is solar. The red, respectively the blue part of the track correspond to the WNL and WNE phases as defined by global properties of the stellar models (see text for the description of the letters). The model is from Barblan (private communication).}
\label{dhrcasa}
\end{center}
\end{figure}

Mass loss can also be triggered by mass transfer in a close binary system. This channel may be important to produce WR stars from lower initial masses or from lower metallicity regions for which the stellar winds alone are too weak to allow the formation of WR stars. An example of such an evolution for a 20 M$_\odot$ plus a 15 M$_\odot$ star is shown in Fig.~\ref{dhrcasa}. Only the evolution of the primary is indicated. 
In the present calculations we did not account for rotation.  At least for the primary this should not be a too severe problem, because after synchronization, the 20 M$_\odot$ model would have
a surface velocity of about 130 km s$^{-1}$ which is a quite moderate velocity. Of course this assumes that the synchronization time is quite short  \citep{deMink09a, Song2015} and does not imply any extra-mixing. The case of the secondary is however quite a different story since it may be spun up by accretion of matter coming from the primary.

In Fig.~\ref{dhrcasa},
point A is the ZAMS. At point B, the radius of the primary reaches for the first time the Roche limit. The mass transfer episode 
lasts 2.3 My, and occurs between points B and G . Note that the strong mass losses occurs in two main events.
The first occurs between points C and D (loss of a little more than 10 M$_\odot$ in a time of about 0.5 My, thus a time-averaged mass loss rate of 2 10$^{-5}$ M$_\odot$ y$^{-1}$). The second
event occurs between points E and G (loss of 3.6 M$_\odot$ in about 0.1 My,  thus a time-averaged mass loss rate of 3.6 10$^{-5}$ M$_\odot$ y$^{-1}$). During these mass transfer events, the mass loss rates due to
Roche Lobe Overflow can be up to 56 times stronger than the wind mass loss rates. The strong mass loss due to the first event makes the track to decrease in luminosity (evolution from C to D in 0.5 My).
At the end of the core hydrogen burning the star contracts and the evolution goes from D to E and x in about 1.2 My. The point x corresponds to the end of the core H-burning phase.
After the core H-burning phase, the core contracts, the envelope expands by mirror effects (evolution from x to F in about 1 My). 
The core helium burning begins in y, still in a mass transfer episode. In y, the star can be considered as
a Wolf Rayet star of the WNL type according to the usual global criteria\footnote{One considers a star to enter
the WR phase when its mass fraction of hydrogen at the surface becomes inferior to 0.3 and the logarithm of the effective temperature is higher than 4.0.}. Due to strong mass losses, which reduce
the H-rich envelope, the star evolves
in the blue region of the HRD.
From G to H, the luminosity decreases again a lot. 
During that phase, the total mass of the primary decreases by
only 0.27 M$_\odot$, but it removes H-rich layers and the mass of hydrogen has an important impact on the radius of the star \citep[see e.g. Fig.~5 and 10 in respectively][]{Groh2013, Meynet2015}.
When the core, at the end of the core He-burning phase begins to contract, the envelope expands (evolution from H to z). In z the star enters into the WNE phase according to the global criteria\footnote{WNE are WR stars defined by the absence of hydrogen at the surface and the usual signatures of CNO processed matter like a high nitrogen abundance.}.
The evolution from z to J corresponds to the contraction phase between the end of the core helium 
burning phase and the beginning of the core carbon phase. At J, the star has reached its final
position in the HRD. 

The close binary evolution described above can therefore produce WR stars. Starting from larger orbital periods shifts the mass transfer at later time, may not produce WR stars and may give birth to redder positions for
the progenitors of the core collapse supernova. We obtain that the upper limit for obtaining WR stars for systems composed of a 20 and a 15 M$_\odot$ is around 6.2 days. All systems with shorter orbital period are predicted to
produce some WR stars at a given point. 

The duration of the WR phase shown in Fig.~\ref{dhrcasa} is a little longer than 1 My, thus quite significant. For comparison, a rotating single 60 M$_\odot$ model at solar metallicity is predicted to be a WR star during 0.9 My \citep{Georgy2012}.
While this kind of evolution might indeed occur in nature, there are a few arguments indicating that it
is probably not too frequent. Let us just mention three of them below.
%\begin{itemize}
%\item
First, In case such an evolution would be common enough, one should find  single-aged massive star populations where both
red supergiants, originating from single or wide binary systems, and WR stars, coming from close binary systems, are observed. At the moment, the stellar clusters at the centre of the Galaxy \citep{Liermann2012} and Westerlund 1 \citep{Clark2005} present evidences for hosting both WR stars and red supergiants.  In general however this is not the case. 
%The fact that the clusters, showing evidences for the presence of both populations, are quite rich clusters may indicate that the frequency of close binary evolution, as sketched in Fig.~\ref{dhrcasa}, is infrequent. It can only be seen when large populations are available.
%\item 
Second, such an evolution would produce low luminous WNL and WNE stars. At the present time, the observed lowest luminosities for these two types of stars are respectively 5.25 and 5.3 in Log L/L$_\odot$, while here we would
have WN and WNE stars with luminosities below 4.9. A way out of this dilemma is either that the frequency of such an evolution is quite small, or that the WR star is hidden in the light of its more luminous companion.
This last point can be checked, and will be discussed in a future paper (Barblan et al. in preparation).
%\item
Third, such an evolution would also produce blue supergiants (actually not the evolution presented in Fig.~\ref{dhrcasa} but close binary evolution with longer orbital periods). Blue supergiants, after the mass transfer, will show a much smaller surface gravity than those coming from stages before the mass transfer. One will have therefore that the post mass transfer blue supergiants will populate a region of the flux weighted gravity luminosity relation that is not compatible with the present day observations \citep{MeynetKud2015}. 
%\item 
%The above arguments needs more study to be more firmly established but at least they give some
%hints on how such evolutions can be tested with observations. 
Another and final point we would like to mention here
is the following:
interestingly, while the close binary scenario can produce low luminous WN stars, it fails in producing low luminous WC stars (at least from the evolution of the primary). According to the observed sample by \citet{Sander2012}, the lowest luminous WC stars have luminosities
around 4.95. The end point of the evolution shown in Fig.~\ref{dhrcasa} is still far from being a WC stars. It would still need to lose 1.5 M$_\odot$. Thus unless there is strong underestimates of the mass losses, there is no chance
to produce low luminous WC stars through this channel for the mass range considered here. This leaves open the question of the origin of these low luminous WC stars. 
%\item 
%\end{itemize}

%\begin{figure}[H]
%\begin{center}
%\includegraphics[width=\columnwidth]{Vsurf_gsurf}
%\caption{
%\label{over}}
%\end{center}
%\end{figure}

%\section{Core overshooting}

%\section{Core magnetic braking and LSGR}

\bibliographystyle{aa} % style aa.bst
\bibliography{myarticle}

\end{multicols}

%\begin{figure}[H]
%\begin{center}
%\includegraphics[width=\columnwidth]{HRD}
%\caption{
%\label{hrd}}
%\end{center}
%\end{figure}

%% file: masterfile.bbl
\begin{thebibliography}{26}
\expandafter\ifx\csname natexlab\endcsname\relax\def\natexlab#1{#1}\fi

\bibitem[{{Aerts}(2008)}]{Aerts2008}
{Aerts}, C. 2008, in IAU Symposium, Vol. 250, IAU Symposium, ed. F.~{Bresolin},
  P.~A. {Crowther}, \& J.~{Puls}, 237--244

\bibitem[{{Brott} {et~al.}(2011){Brott}, {de Mink}, {Cantiello}, {Langer}, {de
  Koter}, {Evans}, {Hunter}, {Trundle}, \& {Vink}}]{Brott2011}
{Brott}, I., {de Mink}, S.~E., {Cantiello}, M., {et~al.} 2011, \aap, 530, A115

\bibitem[{{Clark} {et~al.}(2005){Clark}, {Negueruela}, {Crowther}, \&
  {Goodwin}}]{Clark2005}
{Clark}, J.~S., {Negueruela}, I., {Crowther}, P.~A., \& {Goodwin}, S.~P. 2005,
  \aap, 434, 949

\bibitem[{{de Mink} {et~al.}(2009){de Mink}, {Cantiello}, {Langer}, {Pols},
  {Brott}, \& {Yoon}}]{deMink09a}
{de Mink}, S.~E., {Cantiello}, M., {Langer}, N., {et~al.} 2009, \aap, 497, 243

\bibitem[{{Ekstr{\"o}m} {et~al.}(2012){Ekstr{\"o}m}, {Georgy}, {Eggenberger},
  {Meynet}, {Mowlavi}, {Wyttenbach}, {Granada}, {Decressin}, {Hirschi},
  {Frischknecht}, {Charbonnel}, \& {Maeder}}]{ekstrom12}
{Ekstr{\"o}m}, S., {Georgy}, C., {Eggenberger}, P., {et~al.} 2012, \aap, 537,
  A146

\bibitem[{{Fuller} {et~al.}(2015){Fuller}, {Cantiello}, {Lecoanet}, \&
  {Quataert}}]{Fuller2015}
{Fuller}, J., {Cantiello}, M., {Lecoanet}, D., \& {Quataert}, E. 2015, \apj,
  810, 101

\bibitem[{{Georgy} {et~al.}(2013){Georgy}, {Ekstr{\"o}m}, {Granada}, {Meynet},
  {Mowlavi}, {Eggenberger}, \& {Maeder}}]{Georgy2013}
{Georgy}, C., {Ekstr{\"o}m}, S., {Granada}, A., {et~al.} 2013, \aap, 553, A24

\bibitem[{{Georgy} {et~al.}(2012){Georgy}, {Ekstr{\"o}m}, {Meynet}, {Massey},
  {Levesque}, {Hirschi}, {Eggenberger}, \& {Maeder}}]{Georgy2012}
{Georgy}, C., {Ekstr{\"o}m}, S., {Meynet}, G., {et~al.} 2012, \aap, 542, A29

\bibitem[{{Groh} {et~al.}(2014){Groh}, {Meynet}, {Ekstr{\"o}m}, \&
  {Georgy}}]{Groh2014}
{Groh}, J.~H., {Meynet}, G., {Ekstr{\"o}m}, S., \& {Georgy}, C. 2014, \aap,
  564, A30

\bibitem[{{Groh} {et~al.}(2013){Groh}, {Meynet}, {Georgy}, \&
  {Ekstr{\"o}m}}]{Groh2013}
{Groh}, J.~H., {Meynet}, G., {Georgy}, C., \& {Ekstr{\"o}m}, S. 2013, \aap,
  558, A131

\bibitem[{{Heger} {et~al.}(2005){Heger}, {Woosley}, \& {Spruit}}]{Heger2005}
{Heger}, A., {Woosley}, S.~E., \& {Spruit}, H.~C. 2005, \apj, 626, 350

\bibitem[{{Liermann} {et~al.}(2012){Liermann}, {Hamann}, \&
  {Oskinova}}]{Liermann2012}
{Liermann}, A., {Hamann}, W.-R., \& {Oskinova}, L.~M. 2012, \aap, 540, A14

\bibitem[{{Maeder} \& {Meynet}(2005)}]{MMIII2005}
{Maeder}, A. \& {Meynet}, G. 2005, \aap, 440, 1041

\bibitem[{{Maeder} \& {Meynet}(2012)}]{MM2012}
{Maeder}, A. \& {Meynet}, G. 2012, Reviews of Modern Physics, 84, 25

\bibitem[{{Maeder} \& {Meynet}(2014)}]{MMIV2014}
{Maeder}, A. \& {Meynet}, G. 2014, \apj, 793, 123

\bibitem[{{Meynet} {et~al.}(2015{\natexlab{a}}){Meynet}, {Chomienne},
  {Ekstr{\"o}m}, {Georgy}, {Granada}, {Groh}, {Maeder}, {Eggenberger},
  {Levesque}, \& {Massey}}]{Meynet2015}
{Meynet}, G., {Chomienne}, V., {Ekstr{\"o}m}, S., {et~al.} 2015{\natexlab{a}},
  \aap, 575, A60

\bibitem[{{Meynet} {et~al.}(2015{\natexlab{b}}){Meynet}, {Kudritzki}, \&
  {Georgy}}]{MeynetKud2015}
{Meynet}, G., {Kudritzki}, R.-P., \& {Georgy}, C. 2015{\natexlab{b}}, \aap,
  581, A36

\bibitem[{{Meynet} \& {Maeder}(2005)}]{MMWR2005}
{Meynet}, G. \& {Maeder}, A. 2005, \aap, 429, 581

\bibitem[{{Sander} {et~al.}(2012){Sander}, {Hamann}, \& {Todt}}]{Sander2012}
{Sander}, A., {Hamann}, W.-R., \& {Todt}, H. 2012, \aap, 540, A144

\bibitem[{{Song} {et~al.}(2013){Song}, {Maeder}, {Meynet}, {Huang},
  {Ekstr{\"o}m}, \& {Granada}}]{Song2013}
{Song}, H.~F., {Maeder}, A., {Meynet}, G., {et~al.} 2013, \aap, 556, A100

\bibitem[{{Song} {et~al.}(2016){Song}, {Meynet}, {Maeder}, {Ekstr{\"o}m}, \&
  {Eggenberger}}]{Song2015}
{Song}, H.~F., {Meynet}, G., {Maeder}, A., {Ekstr{\"o}m}, S., \& {Eggenberger},
  P. 2016, \aap, 585, A120

\bibitem[{{Spruit}(2002)}]{Spruit02}
{Spruit}, H.~C. 2002, \aap, 381, 923

\bibitem[{{Suijs} {et~al.}(2008){Suijs}, {Langer}, {Poelarends}, {Yoon},
  {Heger}, \& {Herwig}}]{Suijs2008}
{Suijs}, M.~P.~L., {Langer}, N., {Poelarends}, A.-J., {et~al.} 2008, \aap, 481,
  L87

\bibitem[{{Sz{\'e}csi} {et~al.}(2015){Sz{\'e}csi}, {Langer}, {Yoon}, {Sanyal},
  {de Mink}, {Evans}, \& {Dermine}}]{Szecsi2015}
{Sz{\'e}csi}, D., {Langer}, N., {Yoon}, S.-C., {et~al.} 2015, \aap, 581, A15

\bibitem[{{Zahn}(1992)}]{Zahn1992}
{Zahn}, J.-P. 1992, \aap, 265, 115

\bibitem[{{Zahn} {et~al.}(2007){Zahn}, {Brun}, \& {Mathis}}]{Zahn07}
{Zahn}, J.-P., {Brun}, A.~S., \& {Mathis}, S. 2007, \aap, 474, 145

\end{thebibliography}
